\documentclass[preprint,12pt]{elsarticle}

\usepackage{graphicx}
\usepackage{amssymb}
\usepackage{amsmath}

\usepackage{lineno}

\usepackage[utf8]{inputenc}
\usepackage[T1]{fontenc}
\usepackage{float}
\usepackage{natbib}
\biboptions{square}

\journal{Micron}

\begin{document}

\begin{frontmatter}


\title{Learning-based Defect Recognition for Quasi-Periodic Microscope Images}





\author[1,2]{Nik Dennler\fnref{*}}
\author[1]{Antonio Foncubierta-Rodriguez}
\author[3]{Titus Neupert}
\author[1]{Marilyne Sousa\fnref{*}}
\address[1]{IBM Research Europe - Zurich, Rüschlikon, 8803, Switzerland}
\address[2]{University of Zurich and ETH Zurich, Institute of Neuroinformatics, Zurich, 8057, Switzerland}
\address[3]{University of Zurich, Department of Physics, Zurich, 8057, Switzerland}
\fntext[*]{Corresponding authors. Email address: nik.dennler@uzh.ch; sou@zurich.ibm.com}

\begin{abstract}
Controlling crystalline material defects is crucial, as they affect properties of the material that may be detrimental or beneficial for the final performance of a device. Defect analysis on the sub-nanometer scale is enabled by high-resolution (scanning) transmission electron microscopy [HR(S)TEM], where the identification of defects is currently carried out based on human expertise. However, the process is tedious, highly time consuming and, in some cases, yields ambiguous results. Here we propose a semi-supervised machine learning method that assists in the detection of lattice defects from atomic resolution microscope images. It involves a convolutional neural network that classifies image patches as defective or non-defective, a graph-based heuristic that chooses one non-defective patch as a model, and finally an automatically generated convolutional filter bank, which highlights symmetry breaking such as stacking faults, twin defects and grain boundaries. Additionally, we suggest a variance filter to segment amorphous regions and beam defects. The algorithm is tested on III-V/Si crystalline materials and successfully evaluated against different metrics, showing promising results even for extremely small training data sets. By combining the data-driven classification generality, robustness and speed of deep learning with the effectiveness of image filters in segmenting faulty symmetry arrangements, we provide a valuable open-source tool to the microscopist community that can streamline future HR(S)TEM analyses of crystalline materials. 
\end{abstract}


\end{frontmatter}



\section{Introduction}
\begin{figure}[h]
    \centering
    \includegraphics[width=0.999\linewidth]{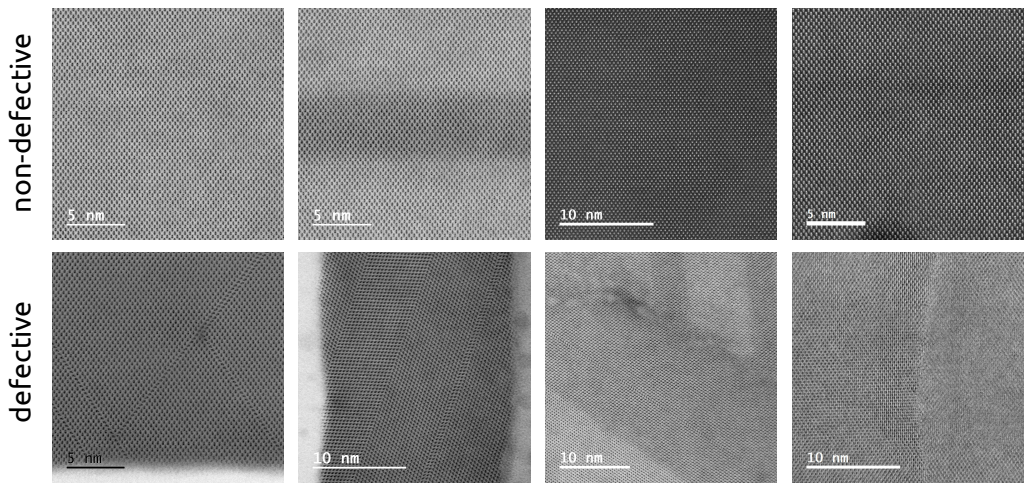}
    \caption{Examples of HRSTEM images where no defects are present (top) and where amorphous regions, twins, stacking faults and grain boundaries are present (bottom).}
    \label{fig:samples}
\end{figure}

Over the last decades we have seen a tremendous growth in the development and use of semiconductor electronics. Silicon-based CMOS technology has been the state-of-the-art for many years, although computationally intensive algorithms and novel computer architectures push the boundaries to the limits. For conducting research on post-silicon materials and determining their optimal growth conditions, the ability to correctly evaluate crystalline defects is crucial, as those often impact properties that are detrimental or enhancing to the final device performance \citep{Ehrhart}. High Resolution (Scanning) Transmission Electron Microscopy [HR(S)TEM] provides sub-nanometer resolution, and thus allows for directly observing the symmetry arrangements of the atoms \citep{HRTEM}. Those reveal if the lattice is perfectly periodic or inherits defects, which are defined as non-periodic structural features such as stacking faults, twin defects, grain boundaries or amorphous regions. Some examples are shown in Figure \ref{fig:samples}. The analysis requires substantial knowledge and often tedious manual work. Furthermore, the results are naturally biased by the person performing the analysis, which can lead to ambiguities in the detected features \citep{Li2018}. The microscopists workflow would benefit from an automated defect detection assistance system that increases the robustness and reproducibility as well as reduces the workload. Additionally it would make the imaging technique more accessible to scientists that are not experts in the field. 

The inspection of material defects from images has a long history from various fields, particularly because it is often an essential step in quality control \citep{2011_Henry_FabricReview}. Previously proposed approaches to automatically capture the symmetry breaking characteristics of material defects can be divided into statistical (auto-correlation functions \citep{18}, co-occurrence matrices \citep{9}, mathematical morphology \citep{15}, fractal methods \citep{65}), spectral (Fourier transform \citep{18}, Wavelet transform \citep{40}, Gabor transform \citep{49}, frequency-filtering \citep{95}), model-based (autoregressive models \citep{100}, template-matching \citep{16}, Markov-random-fields \citep{109}), information theoretical (network analysis/community detection \citep{Ronhovde2012}\citep{Hu_2012}) and learning-based methods (Support-Vector-Machines \citep{SVM}, neural networks \citep{117}). Recent advances in supervised machine learning, i.e., the development of convolutional neural networks \citep{alexnet} (CNNs), offer data-driven solutions to the more general problem of semantic image segmentation \citep{fcnn}, which is the pixel-wise image classification based on learned representations. Those methods have been successfully applied to image analysis tasks involved in plant segmentation \citep{Kattenborn2019}, autonomous driving \citep{semantic_driving}, precision agriculture \citep{semantic_agriculture}, facial segmentation \citep{semantic_face}, medical diagnostics \citep{semantic_diagnostics}, cell biology \citep{2015_Ronneberger_UNet} and more recently material science \citep{Roberts2019_TEMSegmentation}. For CNNs, the performance is best in situations where the to-be-segmented features either stand out from the background through local intensity contrasts or through a distinction in the overlaying texture, and where there is an abundance of training images available.

The challenge of detecting faulty symmetry arrangements in an atomic-resolved crystalline structure is two-fold: First, the notion of what is `defective' is not scale invariant. Consider for example a stacking fault region, which might appear as non-defective if a small enough window-of-sight is chosen but defective if a larger area is considered. A purely local search might yield ambiguous results, thus one requires either information from the global image context or prior knowledge about the materials intrinsic symmetry. Second, some defects might emerge gradually or at continuous levels of severity. Hereby, dividing the image in distinct classes, as semantic segmentation suggests, is bound to result in a loss of information. In this situation, CNN-based semantic segmentation methods would require receptive fields larger than the defects \citep{zeiler2014}, and simultaneously the ability to distinguish small, high-frequency patterns from one another. Both can be achieved by deep and broad networks and large filter kernels, which results in the requirement of very high number of trainable parameters and thus in large training data sets \citep{yu2015lsun} with exhaustive, unambiguous manual annotations \citep{2015_Ronneberger_UNet}. Contrary, filter-based methods have been shown to tackle the unsupervised detection of gradually emerging defects exceptionally well \citep{filter_review}, although often requiring the filter masks either being hand-designed \citep{GaborEdgeDetection} or constructed from known images \citep{templatematching} or parts thereof \citep{AdeEigenfilter}. This makes the methods indeed appropriate for analysing known materials, but rather unsuitable for the analysis of novel materials. A recently proposed method that combines spectral analysis with machine learning to automatically analyse microscopic images \citep{JANY2020102800} appears to be highly effective in decomposing large crystal domains, but has not been demonstrated to detecting symmetry defects on the atomic scale.  

In this work we propose a semi-supervised learning method for the assistance in the detection of lattice defects in sub-nanometer resolution microscope images, which is robust to dealing with ambiguities in the ground-truth training labels and reliable in yielding reproducible results even when trained on a few images only. The system highlights lattice symmetry defects according to their severity and segments amorphous regions. By using a binary convolutional network and automatically generated filter banks, we combine the data-driven classification generality, robustness and speed of deep learning with the effectiveness of filters in detecting faulty symmetries based on image statistics. An extensive image augmentation procedure and transfer learning is used to cope with very small data sets when training the network. The algorithm is tested on HRSTEM images of III-V/Si crystalline materials and successfully evaluated against different metrics.


\section{Materials and Methods}
\begin{figure}[t]
    \centering
    \includegraphics[width=0.99\linewidth]{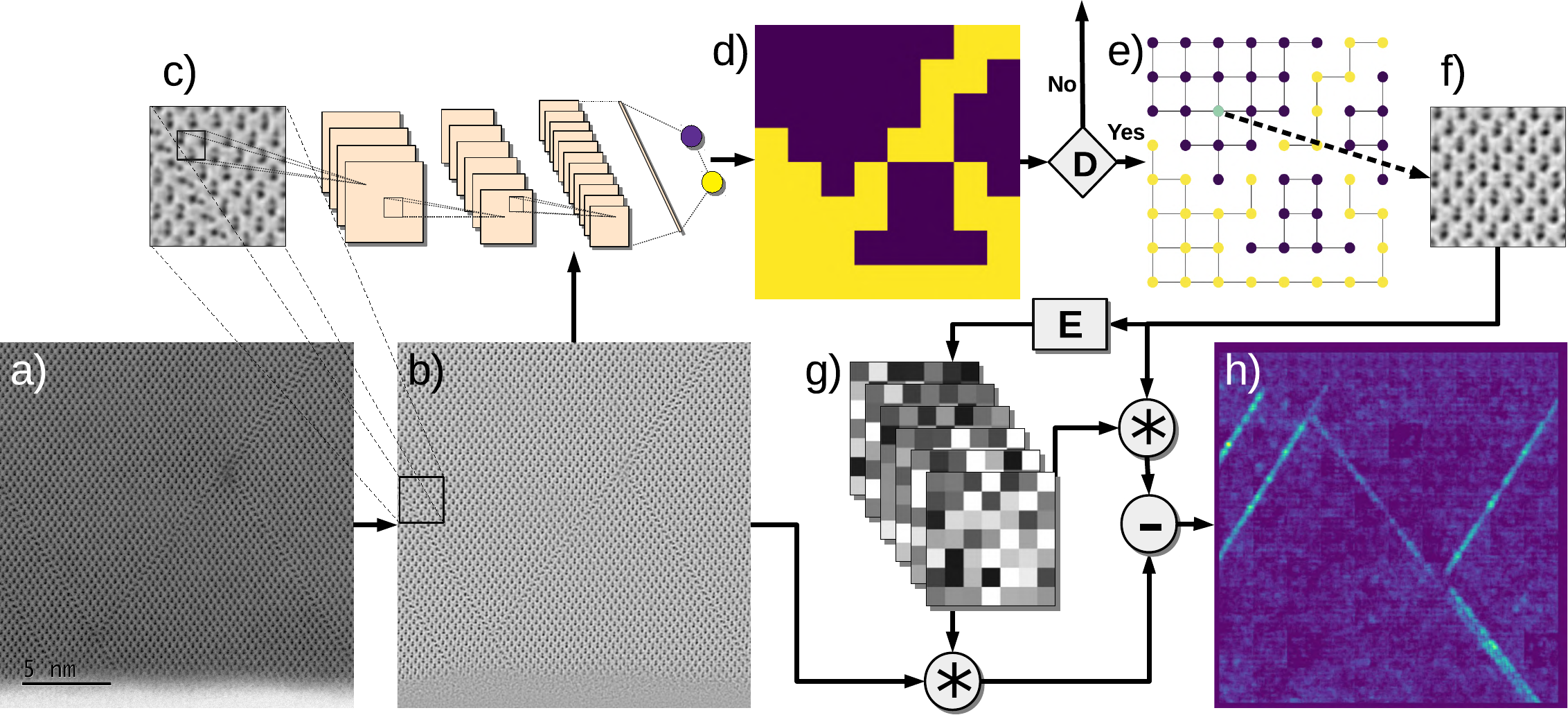}
    \caption{Graphical summary of the method. a) Raw high resolution scanning transmission electron microscopy (HRSTEM) image. b) Image after preprocessing with Laplacian-of-Gaussian filter and Gaussian filter. c) Convolutional neural network, classifying image patches as either non-defective (purple) or defective (yellow). d) Binary prediction map. If no defective patches present, classify image as non-defective and abort. Else, compute 4-connected graph representation e). The green node represents the patch optimal with respect to a weighted-neighborhood norm, giving rise to f), the optimal filter model patch. g) Resulting \textit{Eigenfilters}, which are convolved with the preprocessed image and the optimal patch respectively. The Mahalagonis distance between those two outputs results in h), the final defect segmentation.}
    \label{fig:method}
\end{figure}

The pipeline that we build up is as follows: A convolutional neural network is used for classifying image patches into defective or non-defective. Based on a graph heuristic, one of the non-defective patches is chosen as a model. Subsequently, the image is filtered with an automatically designed mask that is optimal with respect to the symmetry of the non-defective model patch and thus highlights regions affected by stacking faults, twin defects and grain boundaries. The workflow of the method is illustrated in Figure \ref{fig:method}. Additionally, the images are independently filtered with a variance mask, which highlights amorphous regions or beam defects. This section explains the key steps in building up the algorithmic pipeline.

\subsection{Data Acquisition and Image Labeling}
Digital images of size $1024px \times 1024px$  have been acquired with a double spherical aberration corrected high resolution analytical (S)TEM of type \textit{JEOL ARM 200F}, which can achieve resolutions down to 80 picometers . A data set of various materials of cubic-crystalline structure was assembled, resulting in 13 high resolution images containing defects and 28 images not containing defects. The defects in the image were segmented by hand by a human expert. Subsequently, a routine transformed the segmented images into binary masks, which are then used as pixel-wise labels for the supervised-learning process. 

\subsection{Image Preprocessing}
In order to generalize for various illumination and saturation settings, some preprocessing steps are necessary. By filtering the image with a Laplacian of Gaussians \citep{Marr1982} of kernel size $19px \times 19px$, the low frequency changes in intensity are reduced while the lattice structure is preserved. Subsequently, the image is filtered with a Gaussian of kernel size $5px \times5px$, which represses the high frequency noise while smoothing the image. 

\subsection{Data Separation and Augmentation}
Of the 13 defect-containing and labeled images, we randomly select 10 for training and 3 for testing the classifier. This process is repeated six times to ensure generality. 
For the training, each image of size $1024px \times 1024px$ is randomly cropped into patches of size $128px \times 128px$. One way to tackle the risk of overfitting due to the considerably small training data set is to produce more training samples \citep{augmentation}, which we accomplish by augmenting the original images: Performing random rotation and reflection transformations results in a orientation invariant data set. Additionally, in order to simulate a balance between bright-field and dark-field images, the intensity values of whole patches is inverted at random. Each patch is then checked for the labeled defective area and categorized into \textit{non-defective} (relative defective area $< 1\%$), \textit{defective} (relative defective area $> 10\%$) or \textit{ambivalent} ($1\% <$ relative defective area $< 10\%$), where the latter category is not considered for training. In this fashion, we acquire a relatively balanced training data set of around 2,500 image patches per class. For testing, the 10 training images, the 3 test images as well as the 28 non-defective images were cropped in non-overlapping patches of size $128px \times 128px$ with no augmentation applied.

\subsection{Binary Classification via Convolutional Neural Network}
For the classification of the patches into \textit{defective} and \textit{non-defective}, the convolutional neural network architecture \textit{VGG16} \citep{vgg16} is used, which combines 13 convolutional layers using filters of size $3\times3$ pixels, with a number of pooling layers and fully connected layers. To account for the smallness of the data set, transfer learning is leveraged by using synaptic weights that have been pre-trained on the ImageNet data set \citep{imagenet}. For training, the last layer is clipped and replaced with three subsequent fully connected layers with 1024, 1024 and 512 neurons, respectively (we use dropout $d=0.5$ during training). The final output layer has 2 neurons. The complete architecture is shown in Appendix Table~\ref{tab:architecture}. All network parameters are trained with the Adam optimizer \citep{adam} using the categorical crossentropy-loss. A hyper-parameter grid search leaves us with a batch size of 8 patches and an initial learning rate of $l = 10^{-5}$. For building the network model, as well as for training and inference, the Python implementation of the deep learning framework TensorFlow \citep{tensorflow} is used. 

\subsection{Best-patch Heuristics}
Aiming at constructing the optimal filter mask representing the symmetric structure, one wants to select the best model patch from the ones classified as non-defective. Considering the possibility of unforeseen effects close to the defects or to the image border, the desired patch is located in the center of a locally connected neighborhood of defect-free patches. To extract such a patch, we propose to transform the prediction map to a 4-connected graph, from which the one node can be chosen that maximizes a weighted-nearest-neighbor norm defined by
	\begin{equation*}
		W(node) = \sum_{i=1}^k N_i(node) \frac{8^{1-i}}{i!},
	\end{equation*}
where $N_i(node)$ denotes the number of $i$-th order neighbors and $k$ the maximal neighborhood-order. It is easily shown that this norm has the desirable property of guaranteeing that, assuming finite and 4-connected graphs, the minimal contribution of the  $i$-th order neighbors is strictly larger than the maximal possible contribution of all subsequent order neighbors, i.e., the $i$-th order neighborhood contribution cannot be overpowered by all following orders. 

\subsection{Filter construction}
Based on the learned model patch, two sets of filters are constructed: One that highlights changes in symmetry such as stacking faults, twin defects and grain boundaries and one that highlights regions of low variance, such as amorphous regions and beam defects. 
\paragraph{Symmetry filtering}
For obtaining the optimal symmetry filters, we use a variant of the "\textit{Eigenfilter}" technique proposed by Ade \citep{AdeEigenfilter} and further improved by Dewaele et al. \citep{VGoolEigenfilter}: First, a fixed size square kernel (here $7px$ across) is shifted across the model patch, extracting local neighborhood regions and vectorizing them. From the set of vectors, the variance-covariance matrix is calculated and the ordered set of eigenvectors computed. Each eigenvectors represent a set of convolution filter coefficients that stands orthogonal to the others. If assembled back to the original kernel shape, it allows to extract one key features of the structure. For the defect segmentation, each learned filter kernel is convolved with the model patch and an image patch of the same size respectively. The Mahalanobis distance \citep{mahalanobis} between those outputs is calculated, which finally is a measure of defectiveness in the image patch. 

\paragraph{Variance filtering}
To extract local variance values, a square uniform filter of fixed kernel size (here $20px$ across) is convolved once with the image and once with the image after all pixel have been squared, yielding the local mean and the local mean squared respectively. The difference between the squared-mean and the mean-squared yields then a pixel-wise map of local variance as follows:

\begin{equation*}
    \text{Var}(image) =  G_{k\times k}*image^2 - (G_{k\times k}*image)^2,
\end{equation*}
where $G_{k\times k}$ is a uniform kernel of size $k$ and the symbol $*$ denotes the convolution. The same filter is applied to the model patch, which yields appropriate lower-bound values for thresholding the variance map. 

\subsection{Evaluation Metrics}
For the evaluation of the binary patch prediction, we use four metrics that are common in image classification tasks. The metrics are all based on the patch-wise confusion matrix, which consists of true positive (TP), true negative (TN), false positive (FP) and false negative (FN). 
The accuracy, $(TP + TN) / (TP + FP + TN + FN)$, gives the percentage of patches correctly predicted by the binary classifier.
The precision, $TP / (TP + FP)$,  is the ratio of the correctly as defective predicted patches among all as defective predicted patches, penalizing the non-defective cases that have been predicted as defective.
Specificity, $TN / (TN + FP)$, denotes the ratio of the patches correctly predicted as non-defective among all non-defective patches, penalizing the non-defective cases that have been predicted as defective.
Finally, sensitivity or recall, $TP / (TP + FN)$, stands for the ratio of the correctly as defective predicted patches among all defective patches, penalizing the defective cases that have been predicted as non-defective.

As a complementary measure for the quality of the final filtering results, we let a human expert judge the resulting defect detection assistance with their evaluation on the original images and rated the assistance as either \textit{Very helpful}, \textit{Potentially helpful} or \textit{Inadequate}.
Here, \textit{Very helpful} designates the cases where all present defects are detected and there are no significant false-positives. \textit{Potentially helpful} indicates that not all defects are detected and / or there are some false positives, but the filtering results still give rise to the structure in a meaningful way, e.g., successfully assisting in distinguishing between symmetry domains. \textit{Inadequate} relates to the faulty segmentation of defects without yielding additional information.

\section{Results and Discussion}
\begin{table}[t]
\centering
\scriptsize
\begin{tabular}{|l|l|l|l|l|}
\hline
\textbf{Evaluation Sets} & \textbf{Accuracy} & \textbf{Precision} & \textbf{Specificity} & \textbf{Recall} \\ \hline

Train & $92.43\% \pm 1.45\%$ & $87.65\% \pm 3.50\%$ & $93.38\% \pm 2.31\%$ & $90.39\% \pm 3.62\%$ \\ \hline
Test & $88.06\% \pm 9.22\%$ & $79.67\% \pm 18.01\%$ & $89.01\% \pm 12.25\%$ & $82.60\% \pm 16.16\%$ \\ \hline
Specificity test & $99.53\% \pm 0.68\%$ & - & $99.53\% \pm 0.68\%$ & - \\ \hline
\end{tabular}
\caption{Model performance based on different evaluation metrics and evaluation sets. Specificity test set contains unseen and non-defective images only. Mean and standard deviation were calculated from the evaluation results of six models that have been trained on different train sets.} 
\label{tab:model_evaluation}
\end{table}

\subsection{Model Evaluation} 
Table \ref{tab:model_evaluation} shows the evaluation results of the neural network classifier. The metrics were applied for each model to its respective training and test set (no augmentation). The edge cases that were excluded during training, e.g., the image patches containing $1\%$--$10\%$ relative defective area, were included for the evaluation as well. Further, the classifier was evaluated on a specificity test set, consisting of 28 microscope images that do not contain defects at all. The mean and standard deviation for each metric were calculated by considering six different training-test splits of the data and their corresponding trained models. 
The small difference between the evaluation scores of training and test set indicates the successful avoidance of strong overfitting of the model to specific samples in the training set. 
Another noteworthy observation is, that the model performs extremely well for unseen non-defective images. From the six models, the one resulting in the best overall recall value was selected as the production model, which is used to predict the best non-defective model patch to design the filters on. 

\begin{figure}[t]
    \centering
    \includegraphics[width=0.99\linewidth]{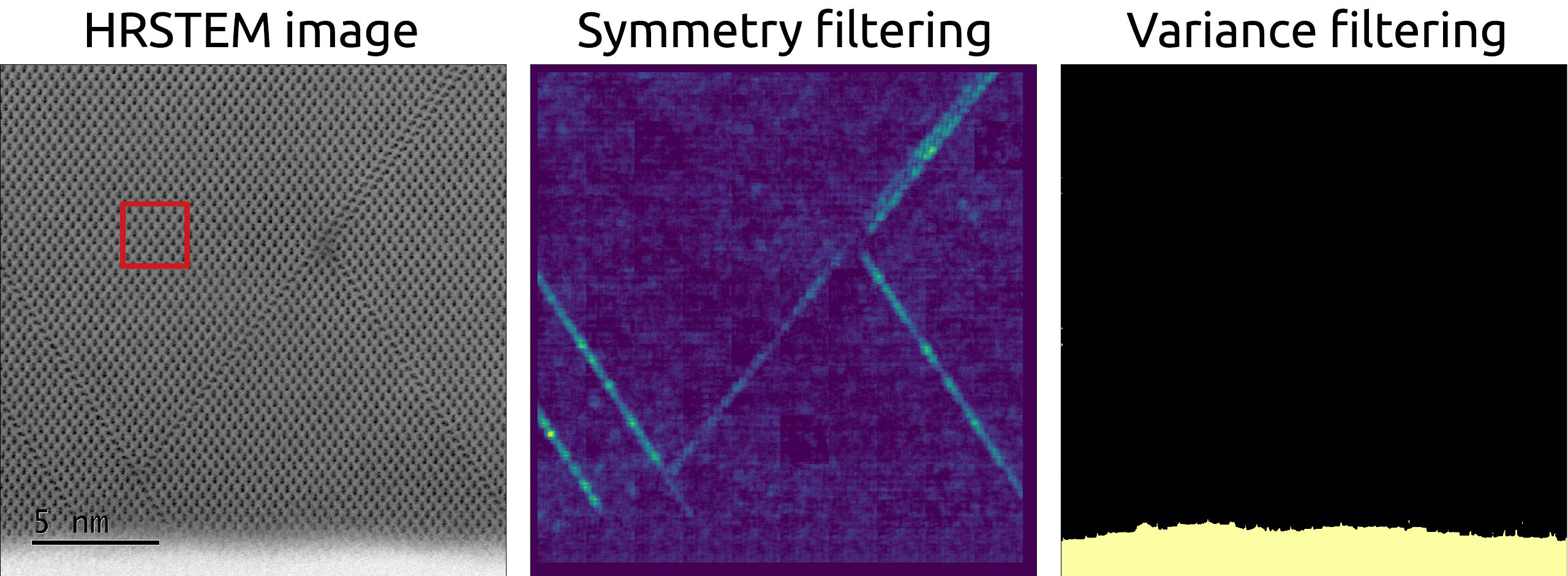}
    \caption{Example highlighting of symmetry defects and segmentation of low variance areas. Left: Raw Scanning Transmission Electron Microscopy (STEM) image with red square indicating the deduced model patch. Center: Symmetry-filtering results, yielding the segmentation of symmetry defects, which in this case arise from multiple stacking faults. The brightness of each pixel indicates the severity of the symmetry defect. Right: Variance-filtering and thresholding results, yielding areas of low local variance, which here are inherited by a continuous amorphous region.}
    \label{fig:results}
\end{figure}

\begin{table}[t]
\centering
\scriptsize
\begin{tabular}{|l|l|l|l|}
\hline
\textbf{Expert Evaluation} & \textbf{Very helpful} & \textbf{Potentially helpful} & \textbf{Inadequate} \\ \hline
Absolute Number & \begin{tabular}[c]{@{}l@{}}9 of 13\end{tabular} & 3 of 13 & 1 of 13 \\ \hline
Percentage & $69.2\%$ & $23.1\%$ & $7.7\%$ \\ \hline
\end{tabular}
\caption{Expert evaluation of the filtering results.}
\label{tab:assistance_evaluation}
\end{table}

\subsection{Filtering Evaluation} 
Figure \ref{fig:results} presents one example of the result obtained by convolving the image with the learned symmetry filter and with a variance filter, where to the latter a subsequent intensity thresholding step is applied. Appendix Figures~\ref{fig:all1} and~ \ref{fig:all2} show a summary of all results. The produced highlighting of symmetry defects and segmentation of high-variance regions are presented to a material characterization expert and rated for their usability. The resulting expert evaluation is presented in Table \ref{tab:assistance_evaluation}.  

\subsection{Computational Time}
The initial training of the VGG16 networks takes 1-2 hours on common GPU machines. The patch-wise prediction and the filtering was done using one i7 CPU core of a standard notebook. The former takes 3-5 seconds, while the latter takes around 2-3 minutes.

\section{Conclusion}
In this work, we presented a novel defect detection assistance system for images of periodic structures, which is based on applying appropriate pre-processing steps, learning the difference from non-defective image patches to defective patches, subsequently finding the best patch using graph heuristics and finally convolving the image with two automatically designed sets of filters. If defects are found, the output of the system is (i) an image with the symmetry faulty region  highlighted, and (ii) an image where the high variance region is segmented. Evaluating on different metrics, we conclude that the classifier yields overall good results and enables the prediction of an adequate model patch. Although the filtering is only applied to a small set of defect-containing images, it is noteworthy than only one time the filtering approach fails completely to assist in detecting anomalies, where in almost $70\%$ of the cases the assistance was rated as very helpful. With additional regards to the examples where the assistance accurately, fully automatically and within seconds predicts the absence of defects for images that do not contain defects, we conclude that experienced and casual users will benefit from a significant reduction of the time and effort spent inspecting materials. 

The most immediate use case of the developed assistance pipeline lays in the HR(S)TEM-based material analysis process where data is scarce and novelty is anticipated, as illustrated through the provided examples. Here, it could also provide reliable and fast analysis of multiple time series images. Moreover, such algorithm could open the HR(S)TEM usage to physicists without much expertise with the instrument. Our workflow can further be applied to a range of characterization techniques beyond HR(S)TEM, where promising candidates might be other atomic imaging techniques, such as scanning electron microscopy, scanning tunneling microscopy and atomic force microscopy. 

In the future, the assistance pipeline could be extended by, for example, the classification of broader types of crystallographic defects and pointing out yet unidentified crystallographic arrangements. The entire code we developed is documented and accessible from our public repository site\footnote{https://github.com/nkdnnlr/TEMDefectClassification}.

\section{Declaration of Competing Interest}
The authors declare no competing financial interests. 


\section{Acknowledgements}
The authors thank S. Mauthe, Y. Baumgartner and H. Schmid for their technical support, P. Staudinger for providing HRSTEM images, K. E. Moselund, R. Hu and C. E. Nauer for fruitful discussions. T.N. acknowledges funding from the European Research Council (ERC) under the European Union’s Horizon 2020 research and innovation program (ERC-StG-Neupert-757867-PARATOP). M.S acknowledges funding from the  DESIGN EID project under the grant agreement no. 860095 under the European Union's Horizon 2020 research and innovation program.

\section{Author contributions statement}
M.S. obtained the microscope images and gathered additional data,  A.R., T.N. and N.D. conceived the experiments, N.D. designed the pre-processing and algorithmic pipeline, N.D. conducted the experiments and analyzed the results, N.D. wrote the manuscript with support from T.N., A.R. and M.S., M.S. supervised the project. All authors reviewed the manuscript.

\bibliographystyle{elsarticle-num-names}
\bibliography{main.bib}






\appendix

\section{All Results}

\begin{figure}[H]
    \centering
    \includegraphics[width=0.5\linewidth]{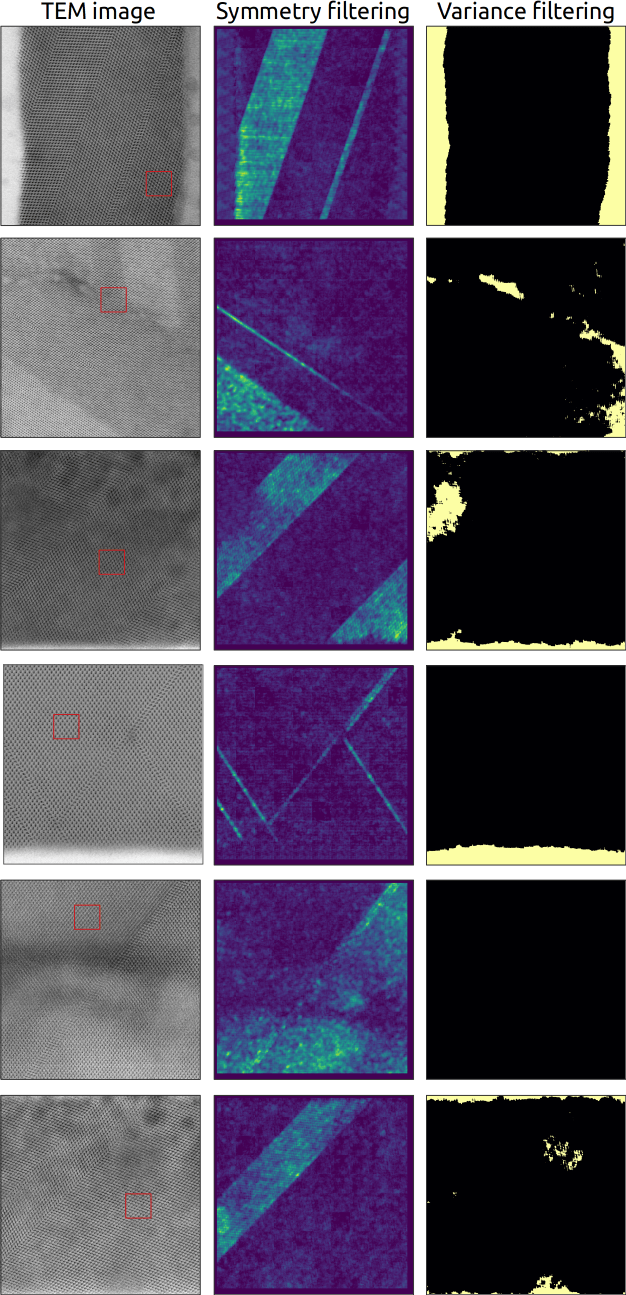}
    \caption{All filtering results. For detailed caption, see Figure \ref{fig:results}}
    \label{fig:all1}
\end{figure}

\begin{figure}[H]
    \centering
    \includegraphics[width=0.5\linewidth]{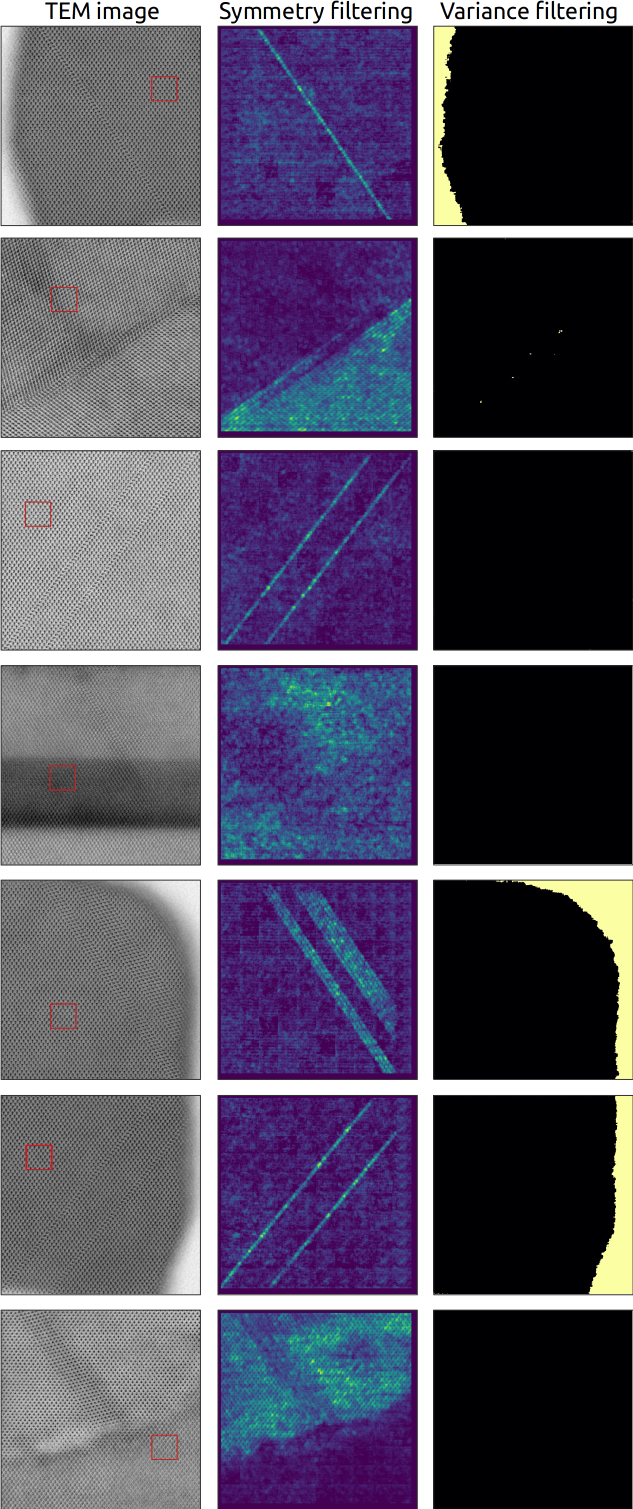}
    \caption{All filtering results. For detailed caption, see Figure \ref{fig:results}}
    \label{fig:all2}
\end{figure}

\section{Network Architecture}

\begin{table}[H]
	\centering
	\begin{tabular}{l|l|ll|l|l}
		\textbf{Layer name} & \textbf{Layer type} & \multicolumn{2}{l|}{\textbf{Output Shape}} & \textbf{Neurons} & \textbf{Parameter} \\ \hline
		input\_1 & InputLayer & 128 & 128 & 3 & 0 \\
		block1\_conv1 & Conv2D & 128 & 128 & 64 & 1792 \\
		block1\_conv2 & Conv2D & 128 & 128 & 64 & 36928 \\
		block1\_pool & MaxPooling2D & 64 & 64 & 64 & 0 \\
		block2\_conv1 & Conv2D & 64 & 64 & 128 & 73856 \\
		block2\_conv2 & Conv2D & 64 & 64 & 128 & 147584 \\
		block2\_pool & MaxPooling2D & 32 & 32 & 128 & 0 \\
		block3\_conv1 & Conv2D & 32 & 32 & 256 & 295168 \\
		block3\_conv2 & Conv2D & 32 & 32 & 256 & 590080 \\
		block3\_conv3 & Conv2D & 32 & 32 & 256 & 590080 \\
		block3\_pool & MaxPooling2D & 16 & 16 & 256 & 0 \\
		block4\_conv1 & Conv2D & 16 & 16 & 512 & 1180160 \\
		block4\_conv2 & Conv2D & 16 & 16 & 512 & 2359808 \\
		block4\_conv3 & Conv2D & 16 & 16 & 512 & 2359808 \\
		block4\_pool & MaxPooling2D & 8 & 8 & 512 & 0 \\
		block5\_conv1 & Conv2D & 8 & 8 & 512 & 2359808 \\
		block5\_conv2 & Conv2D & 8 & 8 & 512 & 2359808 \\
		block5\_conv3 & Conv2D & 8 & 8 & 512 & 2359808 \\
		block5\_pool & MaxPooling2D & 4 & 4 & 512 & 0 \\
		flatten\_1 & Flatten & 1 & 1 & 8192 & 0 \\
		dense\_1 & Dense & 1 & 1 & 1024 & 8389632 \\
		dense\_2 & Dense & 1 & 1 & 1024 & 1049600 \\
		dense\_3 & Dense & 1 & 1 & 512 & 524800 \\
		dense\_4 & Dense & 1 & 1 & 2 & 1026
	\end{tabular}
	\caption{Network architecture, adapted from VGG16}
	\label{tab:architecture}
\end{table}

\end{document}